\documentclass{aa}
\usepackage{graphicx}
\usepackage{txfonts}
\usepackage[numbers]{natbib}

\begin{document}


\title{Retrograde wind accretion --
an alternative mechanism for long spin-period of SFXTs}
\titlerunning{Retrograde Wind Accretion}

\author{J. Wang
          \inst{1}
          \and
          H.-K. Chang
          \inst{1,2}}

\institute{Institute of Astronomy, National Tsing Hua University,
Hsinchu 30013, Taiwan\\
\email{jwang@mx.nthu.edu.tw}
\and Department of Physics, National Tsing Hua University, Hsinchu
30013, Taiwan}


\abstract {A new class of high-mass X-ray binaries (HMXBs) ---
supergiant fast X-ray transients (SFXTs) ---
are discovered by INTEGRAL, which are associated
with OB supergiants and present long spin periods.
Observational evidence indicates that
some accreting neutron stars in HMXBs display accretion reversals.
It has been suggested that the inverted torque can lead to a very slow rotator.
According to three characteristic radii in wind-fed accretion,
we developed a retrograde accretion scenario and divided the accretion phase
into three regimes, to interpret the formation of
the long spin period of SFXTs. The accretion regime in some SFXT systems
can be determined by their spin
and orbital periods.
\keywords{accretion: stellar wind --stars: neutron--
X-rays: binaries} }

\maketitle
\section{Introduction}

The supergiant fast X-ray transient (SFXT) is a new class of massive X-ray binaries
discovered by INTEGRAL observations (Sguera et al. 2005; Negueruela et al. 2006),
which contains a magnetized neutron star that accretes material from
an OB supergiant companion star. They show short and sporadic outbursts (Romano et al. 2007), span 2-5 orders of magnitude in dynamic range (e.g. in't Zand 2005; Leyder et al. 2007), and represent the extreme case
of X-ray variability in high mass X-ray binaries (HMXBs)
with early-type supergiant companions (Masetti et al. 2006b; Pellizza et al. 2006). The quiescent state is characterized by a low luminosity in the range
of $\sim 10^{31}-10^{33}$ ergs/s (e.g. in't Zand 2005; Leyder et al. 2007).

Some SFXTs are believed to be X-ray pulsars with long spin periods,
and their measured pulse periods are located in a wide range, from 4.7 s (AX J1841.0-0536/IGR J18410-0535, Bamba et al. 2001)
to 1240 s (IGR J16418-4532, Walter et al. 2006; Sidoli 2011a).
Generally speaking, a newly born neutron star is presumed
to be a fast rotator with a spin period of a fraction of one second.
If a neutron star is in a binary system, it may experience an ejector state
(the conventional spin-powered energy-loss mechanism) and propeller state
(the interaction between magnetosphere of the neutron star and the stellar wind
of the companion) before accreting material from its optical companion
(for a review see Bhattacharya \& van den Heuvel 1991, and references therein).
During the ejector and propeller regimes, the neutron star spins down and forms a slow rotator
(van den Heuvel 1977; Davies, Fabian \& Pringle 1979; Davies \& Pringle 1981)
at the end of these two phases.
Under certain assumptions (see Urpin, Konenkov \& Geppert 1998a;
Li \& van den Heuvel 1999), the spin period can be limited to a few hundred
seconds ($\sim 500$ s). If the system undergoes a spherical accretion
and three epochs, i.e. the ejector, supersonic propeller,
and subsonic propeller epoch,
the conditions under which a longer period can be achieved
may be relaxed (Ikhsanov 2001; Ikhsanov 2007).

Long-period pulsars are often identified as an accretors
and their periods gradually decrease during the accretion
in the prograde scenario. However, observations indicate
that dynamic torque reversals are present in some wind-fed persistent HMXBs
(Bildsten et al. 1997; Nelson et al. 1997, and references therein).
The spin-down torque due to the accretion reversals has the same order
of magnitude as the spin-up torque, but is opposite in sign (Nelson et al. 1997).
Accretion torque reversals can occur in an inhomogeneous wind (Ruffer 1997, 1999).

In this paper, we suggest that the long spin periods are associated
with a retrograde accretion scenario.
In section 2, we review the theory of wind accretion in SGXBs/HMXBs
and derive the spin evolution and the spin-down timescale
in different accretion regimes.
We apply our model to SFXTs in section 3.
Section 4 contains a short summary.


\section{Retrograde wind accretion}

OB supergiants (Reig 2011) can produce a powerful stellar wind that is driven by radiation,
which is approximately isotropic, that is, the wind material leaves
the surface uniformly in all directions,
with a wind velocity of $v_w \sim 1000-2000$ km~s$^{-1}$.
The OB companions in SFXTs typically have a mass of $M_c \sim 30 M_{\odot}$,
mass loss rate of $\dot{M}_w \sim 10^{-6} M_{\odot}~yr^{-1}$,
and luminosity of $\log(L_c/L_{\odot}) \sim 5-6$.
Here, $M_{\odot}$ is the solar mass, and $L_{\odot}$ is the solar luminosity.

The magnetized neutron star in SGXBs accretes matter from an OB companion via
the intense stellar wind (Verbunt \& van den Heuvel 1995).
We consider the case of a spherically symmetric stellar wind
radiated from the OB companions that is disordered and inhomogeneously and is captured
by relativistic neutron stars. The accreting neutron star can be either
a prograde accretor or a retrograde accretor
(Bildsten et al. 1997; Shakura et al. 2012).
If the neutron star behaves like a retrograde accretor,
it may spin down and rotate with a very long period because
of the inverted accretion torque.

\subsection{Characteristic radii}

In the theory of wind accretion by a magnetized rotating neutron star,
three characteristic radii
(Davidson \& Ostriker 1973; Lamb, Pethick \& Pines 1973;
Stella, White \& Rosner 1986) can be defined as follows.
Only part of the wind material from the OB companion will be captured
by the gravitational field of the neutron star.
The radiated wind flowing toward the neutron star
within a certain radius will be accreted,
whereas the flowing material outside that radius will escape.
This radius, called the accretion radius $r_{acc}$ (Bondi \& Hoyle 1944),
can be defined by noting that material will only be accreted
if it has a kinetic energy lower than the potential energy
in the gravitational field of the neutron star, that is,
\begin{equation}
r_{acc} = 2GM/v_w^2 \sim 10^{10}\frac{M}{M_{\odot}}v_{w,8}^{-2}
\,\mbox{\rm cm}\,\, ,\label{accra}
\end{equation}
where $M$ is the neutron star mass, and $v_{w,8}$ is a velocity in units
of 1000 km~s$^{-1}$ ($v_{w,8}=v_w/(10^8$ cm s$^{-1})$).

When the magnetic pressure of the neutron star can balance
the ram pressure of the accreting material, a magnetosphere forms
and the inflowing matter may stop at this preferred radius
--- the magnetospheric radius $r_{mag}$
(Elsner \& Lamb 1977; Davies \& Pringle 1981). It can be defined by
\begin{equation}
\frac{B(r_{mag})^2}{8\pi}=\rho(r_{mag}) v(r_{mag})^2,
\end{equation}
where $B(r_{mag}) \sim \mu/r_{mag}^3$ is the magnetic field strength at $r_{mag}$, and the
$\rho(r_{mag})$ and $v(r_{mag})$ are wind density and wind velocity at $r_{mag}$, respectively. $\mu$ is the magnetic moment of the neutron star.
For $r_{mag} > r_{acc}$, the inflows cannot experience a significant gravitational field. Therefore, the wind speed at $r_{mag}$ is $v(r_{mag}) = v_w$. Assuming a nonmagnetized spherically symmetric wind (Elsner \& Lamb 1977), the wind density $\rho(r_{mag})$ reads $\rho(r_{mag}) \approx \dot{M}_{w}/4\pi a^2 v_w$ (Davies \& Pringle 1981; Bozzo, Falanga \& Stella 2008),
where $a$ is the orbital separation of the two components in the system
and $a >> r_{mag}$ is assumed.
Therefore, the magnetospheric radius is \begin{equation}
r_{mag} \sim 10^{10}\dot{M}_{w,-6}^{-1/6}v_{w,8}^{-1/6}a_{10}^{1/3}\mu_{33}^{1/3}
\,\mbox{\rm cm}\,\, ,\label{lmagra}
\end{equation}
where $\dot{M}_{w,-6} = \dot{M}_w/(10^{-6}M_{\odot}/{\rm yr})$,
$\mu_{33} = \mu/(10^{33}$ G~cm$^3$), and $a_{10}$ is related to the orbital separation
as $a \sim 10^{12} a_{10}$ cm $\sim 10^{12} P_{b,10}^{2/3} M_{30}^{1/3}$ cm.
$P_{b,10}$ is the orbital period $P_b$ in units of $10$ days,
and $M_{30}$ is the companion mass in units of $30$ solar masses.
When $r_{mag} < r_{acc}$, the gravitational field of the neutron star dominates the falling of wind matter. The wind density and wind velocity at $r_{mag}$ can therefore
be taken as
$\rho(r_{mag}) = \frac{\dot{M}}{v(r_{mag})4\pi r_{mag}^2}$ and $v(r_{mag}) = \sqrt{\frac{2GM}{r_{mag}}}$, respectively. The accretion rate $\dot{M}$ depends on $r_{acc}$ and $a$ according to (Frank, King \& Raine 2002)
\begin{equation}
\dot{M}/\dot{M}_w \sim r_{acc}^2/(4a^2) \sim 10^{-5}\left(\frac{M}{M_{\odot}}\right)^2v_{w,8}^{-4}a_{10}^{-2}\,\, .
\end{equation}
Then the magnetospheric radius is given by
\begin{equation}
r_{mag} \sim 10^{10}\left(\frac{M}{M_{\odot}}\right)^{-5/7}\dot{M}_{w,-6}^{-2/7}v_{w,8}^{8/7}a_{10}^{4/7}\mu_{33}^{4/7}
\,\mbox{\rm cm}\,\, .\label{magra}
\end{equation}

A corotation radius $r_{cor}$ can be defined
because the spin angular velocity ($\Omega_s = 2\pi/P_s$)
of the neutron star is equal to the Keplerian angular velocity
($\Omega_k = \sqrt{\frac{GM}{r_{cor}^3}}$) of matter being accreted,
\begin{equation}
r_{cor} \sim 10^{10}\left(\frac{M}{M_{\odot}}\right)^{\frac{1}{3}}P_{s,3}^{2/3}
\,\mbox{\rm cm}\,\, ,\label{corora}
\end{equation}
where $P_{s,3}$ is the spin period of the neutron star $P_s$
in units of $10^3$ s.
The competition among the three characteristic radii leads
to different accretion regimes, so do the accretion modes
and the behavior of the neutron star.

\subsection{Pure propeller}

In a system in which the magnetospheric radius is larger
than the accretion radius ($r_{mag} > r_{acc}$),
the stellar wind, flowing around the magnetosphere,
may not experience a significant gravitational field
and interacts directly with the magnetosphere.
The neutron star behaves like a pure propeller and
spins down via rotational energy dissipation that results
from the interaction between the magnetosphere and stellar winds,
similar as in radio pulsars with magnetic inhibition
(Stella, White \& Rosner 1986; Bozzo, Falanga \& Stella 2008).
The spin-down torque can be expressed as
(Lynden-Bell \& Pringle 1974; Wang 1981; Lipunov 1992)
\begin{equation}
T_{p,pro} = -\kappa_t \mu^2/r_{mag}^3\,\, ,\label{protor}
\end{equation}
where $\kappa_t \leq 1$ is a dimensionless parameter
of the order of unity (Davies \& Pringle 1981; Lipunov 1992).
The equation governing the spin evolution can be written as
(Davidson \& Ostriker 1973; Ghosh \& Lamb 1978)
\begin{equation}
2\pi I\frac{d}{dt}\frac{1}{P_s} = T \,\, ,\label{spinevo}
\end{equation}
where $I$ is the moment of inertia of the neutron star, and $T$ is the total torque imposed on the neutron star.
Therefore, the spin-down rate reads
\begin{equation}
\dot{P}_{s,pur} \sim 10^{-6}\kappa_t\mu_{33}I^{-1}_{45}P^2_{s,3}a^{-1}_{10}\dot{M}^{1/2}_{w,-6}v^{1/2}_{w,8}~s~s^{-1},
\end{equation}
where $I_{45}$ is the moment of inertia in units of $10^{45}$g~cm$^2$.
Accordingly, if the neutron star spins down to a period of 1000 s, the spin-down timescale is given by
\begin{eqnarray}
\tau_{s,pur} = P_s/\dot{P}_{s,pur} \sim 10^3\kappa^{-1}_t\mu^{-1}_{33} I_{45}P^{-1}_{s,3}a_{10}\dot{M}^{-1/2}_{w,-6}v^{-1/2}_{w,8}~yr.
\end{eqnarray}
The luminosity in this regime results from the shock forming near $r_{mag}$,
\begin{eqnarray}
L_{pur} \sim r^2_{mag}\rho_w v^3_w \sim 10^{32}\dot{M}^{5/6}_{w,-6}v^{11/6}_{w,8}\mu^{1/3}_{33}a^{-5/3}_{10}~ergs~s^{-1}.
\end{eqnarray}

\subsection{Retrograde propeller --- quasi-spherical capture}

If the magnetospheric radius is smaller than the accretion radius,
the stellar wind radiated from the companion can penetrate
through the accretion radius and halt at the magnetopause.
We consider the regime of $r_{mag} > r_{cor}$ in this part.
The inflow in the region between $r_{acc}$ and $r_{mag}$ falls
approximately in a spherical configuration. The neutron star behaves like a supersonic rotator,
i.e., the linear velocity of rotation of magnetosphere is much higher than the velocity of sound in the wind clumps, and ejects some of that wind material out of the accretion radius,
leading to the dissipation of some rotational energy and thus spinning down.
This scenario corresponds to the propeller mechanism
and imposes a spin-down torque on the neutron star.
The torque resulting from the propeller mechanism reads (Lipunov 1992)
\begin{equation}
T_{r,pro} = -\dot{M} v_w^2/\Omega_s\,\, .\label{rprotor}
\end{equation}

In addition, due to the high speed of strong stellar wind,
the accumulation of accreted matter is faster
than the ejection rate by the propeller mechanism.
More and more wind clumps may be deposited outside the magnetosphere (Stella, White \& Rosner 1986). The deposit is disordered by the supersonic rotation of the neutron star, forming turbulent motion around the magnetosphere. The velocity of turbulent motion of accumulated stellar wind at the magnetoaphere boundary is close to the velocity of sound (Lipunov 1992). Because of the disorder of the
stellar wind and the turbulent motion (Davies \& Pringle 1981),
the flowing accumulated material can be either prograde or retrograde
around the magnetosphere.
If the wind flows retrogradely, it will impose an inverted accretion torque
on the neutron star and spin it down.
This accretion torque can be written as (Lipunov 1992)
\begin{equation}
T_{r,acc} = -\dot{M}\Omega_s r_{mag}^2\, \, .\label{racctor}
\end{equation}

Accordingly, the total torque imposed on the neutron star in this regime is
\begin{eqnarray}
T_{p} = T_{r,pro} + T_{r,acc}
= -\dot{M} v_w^2/\Omega_s - \dot{M}\Omega_s r_{mag}^2\,\, . \label{ptotor}
\end{eqnarray}
Substituting Eq. (\ref{ptotor}) into Eq. (\ref{spinevo}),
we can obtain the spin evolutionary law
in the regime of $r_{acc}>r_{mag}>r_{cor}$,
\begin{eqnarray}
\dot{P}_{s,pro}&\sim&10^{-7}I_{45}^{-1}[\dot{M}_{w,-6}v_{w,8}^{-2}a_{10}^{-2}P_{s,3}^3+\left(\frac{M}{M_{\odot}}\right)^{9/7}\nonumber\\
&&\dot{M}_{w,-6}^{5/7}a_{10}^{-10/7}v_{w,8}^{-20/7}\mu_{33}^{4/7}P_{s,3}]~{\rm s}~{\rm s}^{-1}\,\, .
\end{eqnarray}
In this scenario, the neutron star can spin down to 1000 s during a timescale of
\begin{eqnarray}
\tau_{s,pro} &\sim& 10^{4}I_{45}[\dot{M}_{w,-6}v_{w,8}^{-2}a_{10}^{-2}P_{s,3}^2+\left(\frac{M}{M_{\odot}}\right)^{9/7}\nonumber\\
&&\dot{M}_{w,-6}^{5/7}a_{10}^{-10/7}v_{w,8}^{-20/7}\mu_{33}^{4/7}]^{-1}~\mbox{\rm yr}\,\, .
\end{eqnarray}
However, if the stellar wind is not strong enough, the accreting matter is approximately in radial free fall as it approaches the magnetopshere boundary.
The captured material cannot accumulate near the magnetosphere
due to the supersonic rotation of the neutron star.
Consequently, the accretion torque contains only
the spin-down torque imposed by the propeller mechanism.
This alternative scenario is discussed in
Urpin, Geppert \& Konenkov (1998b,c).
It is less relevant for SFXTs, since the OB giant companions emit
strong stellar wind.

In the retrograde propeller phase, both the interaction with magnetic fields by the propeller mechanism
and inverted accretion contribute to the luminosity, which reads
\begin{eqnarray}
L_{pro}&=&\dot{M}v_w^2 + GM\dot{M}/r_{mag} \nonumber\\
&\sim& 10^{32}[10\left(\frac{M}{M_{\odot}}\right)^2\dot{M}_{w,-6}v^{-2}_{w,8}a^{-2}_{10} + \left(\frac{M}{M_{\odot}}\right)^{26/7} \nonumber\\ &&\dot{M}^{9/7}_{w,-6}v^{-36/7}_{w,8}a^{-18/7}_{10}\mu^{-4/7}_{33}]~\mbox{\rm ergs}~{\rm s}^{-1}\,\, .
\end{eqnarray}
In addition, the wind material halts at the magnetopause
and forms a shock near $r_{mag}$. Energy is also released
through the shock, and the corresponding luminosity is given by Eq. (17)
in Bozzo, Falanga \& Stella (2008).

\subsection{Retrograde accretor --- quasi-disk accretion}

When $r_{acc} > r_{mag}$ and $r_{cor} > r_{mag}$,
the disordered stellar wind penetrates
the accretion radius and falls directly toward the magnetosphere.
In this case, the velocity of sound at the magnetosphere boundary is much higher than the linear velocity of the neutron star rotation, and the neutron star is a subsonic rotator.
Owing to the relatively slow rotation of the neutron star,
the rotating magnetosphere cannot eject the wind material. The inflows experience a significant gravitational field and accumulate near the magnetosphere. The accumulated material, with high initial velocity, can form an extended adiabatic, tenuous, and retrogradely rotating atmosphere around the magnetosphere, which impose an inverted accretion torque and spin the neutron star down. In addition, the reversal rotation between the adiabatic atmosphere and the magnetosphere may cause a flip-flop behavior of some clumps in the atmosphere, ejecting a part of material and leading to the loss of thermal energy.

The spin-down torque resulting from the flip-flop ejection
has the same form as Eq. (\ref{protor}) (Davies \& Pringle 1981; Lipunov 1992).
The accretion spin-down torque imposed by the inverted atmosphere has the same expression as Eq. (\ref{racctor}).
Accordingly, the total spin-down torque imposed on the neutron star
in the regime of $r_{acc} > r_{mag}$ and $r_{cor} > r_{mag}$ is
\begin{equation}
T_a = -\kappa_t\frac{\mu^2}{r_{mag}^3} - \dot{M}\Omega_s r_{mag}^2\,\, .
\end{equation}
Substituting the above expression into Eq. (\ref{spinevo}),
we obtain the spin evolution equation,
\begin{eqnarray}
\dot{P}_{s,acc}&\sim&10^{-6}I_{45}^{-1}[\kappa_t\mu_{33}^{2/7}P_{s,3}^2\left(\frac{M}{M_{\odot}}\right)^{15/7}\dot{M}_{w,-6}^{6/7}
v_{w,8}^{-24/7}a_{10}^{-12/7}\nonumber\\
&&+\mu_{33}^{8/7}P_{s,3}\left(\frac{M}{M_{\odot}}\right)^{-14/7}\dot{M}_{w,-6}^{3/7}v_{w,8}^{16/7}a_{10}^{8/7}]~{\rm s}~{\rm s}^{-1}\,\, .
\end{eqnarray}
The spin-down (to a period of 1000 s) timescale reads
\begin{eqnarray}
\tau_{s,acc}&\sim&10^{3}I_{45}[\kappa_t\mu_{33}^{2/7}P_{s,3}\left(\frac{M}{M_{\odot}}\right)^{15/7}\dot{M}_{w,-6}^{6/7}
v_{w,8}^{-24/7}a_{10}^{-12/7}\nonumber\\
&&+\mu_{33}^{8/7}\left(\frac{M}{M_{\odot}}\right)^{-14/7}\dot{M}_{w,-6}^{3/7}v_{w,8}^{16/7}a_{10}^{8/7}]^{-1}~\mbox{\rm yr}\,\, .
\end{eqnarray}
The corresponding luminosity is
\begin{eqnarray}
L_{acc} \sim 10^{31}\left(\frac{M}{M_{\odot}}\right)^{26/7}\dot{M}^{9/7}_{w,-6}v^{-36/7}_{w,8}a^{-18/7}_{10}\mu^{-4/7}_{33}~\mbox{\rm ergs}~{\rm s}^{-1}\,\, .
\end{eqnarray}


\section{Application to SFXTs}

Some SFXTs that contain sporadically accreting neutron stars
are believed to be X-ray pulsars and display very different pulse periods,
from a few seconds (e.g. IGR J16479-4514, $P_s$ = 2.14 s, http://hera.ph1.uni-koeln.de/~heintzma/Integral/SFXT.htm;
IGR J18483-0311, $P_{s}$ = 21 s, Sguera, Hill \& Bird et al. 2007;
IGR J17544-2619, $P_{s}$ = 71.49s, Drave et al. 2012;
AX J1749.1-2733, $P_s$ = 132 s, Karasev et al. 2007, Karasev et al. 2008;
IGR J111215-5952, $P_{s}$ = 187 s, Swank et al. 2007;
IGR J16465-4507, $P_{s}$ = 228 s, Lutovinov et al. 2005)
to a few thousand seconds (e.g. IGR J16418-4532, $P_{s}$ = 1240 s,
Walter et al. 2006, Sidoli et al. 2012;
XTE J1739-302, reported for periodicity in the 1500-2000 s range,
Sguera et al. 2006).
Their orbital periods also span a very wide range,
from 1.4 days to a few hundred days (e.g. Sidoli 2011b, and references therein).

According to the definition of the three aforementioned characteristic radii
(see Eq. (\ref{accra}), Eq. (\ref{magra}) and Eq. (\ref{corora})),
they depend
on some physical properties of the system,
including the neutron star mass and magnetic field strength,
stellar wind velocity, orbital period of the binary system,
the mass loss rate of stellar wind, the mass of the companion, and
the neutron star spin period.
In this work we take the neutron star mass as $1.4 M_{\odot}$
and magnetic field as $10^{12}$ G,
the mass of OB supergiant as $30 M_{\odot}$,
the mass loss rate and velocity of the stellar wind
as $10^{-6} M_{\odot}$ yr$^{-1}$ and 1000 km/s, respectively.
Then the accretion radius is about $10^{10}$ cm, and the corotation radius depends only on the spin period as $r_{cor} \sim P_{s,3}^{2/3}$.
The magnetospheric radius depends also only on the orbital period
according to $r_{mag} \sim P_{b,10}^{2/9}$ in the region of $r_{mag} > r_{acc}$ and $r_{mag} \sim P_{b,10}^{8/21}$ when $r_{mag} < r_{acc}$.
Therefore, we can determine the accretion regime
in which a certain SFXT is located,
according to the spin period and orbital period (see Table 1).

IGR J18483-0311 and IGR J17544-2619 display
not-too-slow rotation and relatively short orbital periods
(for IGR J18483-0311, $P_s$ = 21 s, http://hera.ph1.uni-koeln.de/~heintzma/Integral/SFXT.htm,
$P_{orb}$ = 18 days, Jain et al. 2009;
for IGR J17544-2619, $P_s$ = 71.49 s, Drave et al. 2012,
$P_{orb}$ = 4.9 days, Clark et al. 2010).
From the expressions of $r_{mag}$ and $r_{cor}$
(see Eq. (\ref{magra}) and Eq. (\ref{corora})),
one can see that the relation among these three radii
is $r_{acc} > r_{mag} > r_{cor}$.
This means that IGR J18483-0311-like sources can behave
like a propeller in the retrograde scenario.

Very long spin period SFXTs with narrow orbits,
e.g. IGR J16418-4532 ($P_s$ = 1240 s, Walter et al. 2006, Sidoli et al. 2011a;
$P_{orb}$ = 3.74 days, Levine et al. 2011),
typically have a large corotation radius.
Moreover, the magnetospheric radius is one order of magnitude
smaller than the accretion radius.
Therefore, this class of systems can be an accretor with retrograde accretion. In addition, IGR J16479-4514, with $P_s$ = 2.14 s (http://hera.ph1.uni-koeln.de/~heintzma/Integral/SFXT.htm) and $P_{orb}$ = 3.3 days (Jain et al. 2009), also present a somewhat larger corotation radius. Therefore, it can also behave like an retrograde accretor.

There are several slowly rotating systems with spin periods
of a few hundred seconds and relatively wide orbital separation,
such as IGR J111215-5952 ($P_{s}$ = 187 s, Swank et al. 2007;
$P_{orb}$ = 165 days, Sidoli et al. 2007),
AX J1749.1-2733 ($P_s$ = 132 s, Karasev et al. 2007, Karasev et al. 2008;
$P_{orb}$ = 185.5 days, Zurita \& Chaty 2008),
and IGR J16465-4507
($P_s$ = 228 s, Lutovinov et al. 2005; $P_{orb}$ = 30.32 days,
Clark et al. 2010). As listed in Table 1, the relations among the three aforementioned characteristic radii of these sources are $r_{mag} > r_{acc} > r_{cor}$. Therefore, these three sources can be in the pure propeller regime.

The outbursts of XTE J1739-302 observed by
several observatories show a characteristic timescale
for variability of about 1500-2000 s (Sguera et al. 2006).
This source has an orbital period of 51.47 days (Drave et al. 2010).
Both the corotation radius and magnetospheric radius are slightly larger than the accretion radius (see Table 1 for details). However, due to the uncertainty in the spin period, it is difficult to determine its exact status. More accurate knowledge about the properties of the system is needed to determine whether the system is a pure propeller or a retrograde propeller.

It can be seen that the three characteristic radii in several sources have the same order of magnitude (see Table 1 for details). IGR J16418-4532 shows a dynamical range of two orders of magnitude
and quasi-periodic flares (Sidoli 2011b).
These features were taken to suggest that its X-ray emission is driven
by a transitional accretion regime in a prograde accretion scenario
(Sidoli 2011b). In this sense, we claim that this class of SFXTs frequently experience
a transitional state between the propeller regime and the retrograde accretor phase.

The above distinction depends strongly on the assumption
that these systems have the same properties except
for spin and orbital periods. The particularities for each system should be investigated for
a more detailed study.

\begin{table*}
\caption{Different accretion regimes for 8 SFXTs}
\begin{tabular}{llllll}
\hline\hline
Source \ & \ $P_{s}$($10^3$s) \ &  \
$P_{b}$(10days)  \ & \ $r_{cor}$($10^{10}$cm) \ &
 \ $r_{mag}$($10^{10}$cm) \ & \ Accretion regime \\
\hline
AX J1749.1-2733  \ & \ 0.132  \ &  \
18.55  \ & \ 0.26 \ & \ 1.9 \ & \ pure propeller  \\
IGR J111215-5952 \ & \ 0.187 \ & \
16.5 \ & \ 0.33 \ & \ 1.9 \ & \ pure propeller \\
IGR J16465-4507 \ & \ 0.228 \ & \
3.032 \ & \ 0.37 \ & \ 1.3 \ & \ pure propeller \\
IGR J18483-0311 \ & \ 0.021 \ & \
0.18 \ & \ 0.016 \ & \ 0.52 \ & \ retrograde propeller \\
IGR J17544-2619 \ & \ 0.071 \ & \
0.49 \ & \ 0.17 \ & \ 0.76 \ & \ retrograde propeller \\
IGR J16418-4532 \ & \ 1.24 \ & \
0.374 \ & \ 1.2 \ & \ 0.69 \ & \ retrograde accretor \\
IGR J16479-4514 \ & \ 0.002 \ & \
0.33 \ & \ 0.99 \ & \ 0.65 \ & \ retrograde accretor \\
XTE J1739-302 \ & \ 1.5-2.0 \ & \
5.147 \ & \ 1.3-1.6 \ & \ 1.4 \ & \ uncertain \\
\hline\hline
\end{tabular}
\label{cata}
\end{table*}


\section{Summary}

Neutron stars in a wind-fed accreting system
with disordered and inhomogeneous stellar winds
can be either a prograde or a retrograde accretor.
An accretor may exhibit different behaviors in different accretion regimes,
depending on the competition among the three characteristic radii.

When the magnetospheric radius is located outside the accretion radius, the wind cannot experience significant gravitational focusing and forms a bow shock near the magnetosphere. The interaction between the stellar wind and the magnetosphere contributes to the loss of rotational energy and the spin-down of neutron star. This is the pure propeller regime.
If the accretion radius is larger than the magnetospheric radius,
the wind flowing from OB companions with a high velocity
directly interacts with the magnetosphere and forms a shock
at $r_{mag}$ (Harding \& Leventhal 1992).
In this case, when $r_{mag}$ is larger than the $r_{cor}$,
the neutron star can behave like a retrograde propeller.
The shock, formed near the magnetosphere, flip-flops from side to side and makes the uniform wind disordered and inhomogeneous. Therefore, the neutron star accretes matter
with a relatively high angular velocity, but with opposite signs (Livio 1992),
consistent with the retrograde scenario.
If both the accretion radius and corotation radius are larger
than the magnetospheric radius, the inflowing matter can fall directly
and halt near the magnetopause.
Because of the high falling velocity,
there is no effective cooling mechanism.
Moreover, the dissipation of the rotational energy
also produces additional heat.
The spherically symmetric accreting material
forms a quasi-disk, adiabatic and tenuous atmosphere around the neutron star, which can retrogradely rotate with the magnetosphere and spin the neutron star down. In the case of retrograde scenario, the neutron star can spin down to a period of 1000 s during a timescale of about $10^3$ - $10^4$ yr.


\section*{Acknowledgements}

It is a pleasure to appreciate enlightening suggestions from S. N. Zhang.
This work was supported by the National Science Council
of the Republic of China under grant
NSC 99-2112-M-007 -017 -MY3.

\end{document}